\documentclass[aps,prd,preprint,superscriptaddress]{revtex4}

\usepackage{ulem}
\usepackage{xcolor}
\usepackage[colorlinks=true]{hyperref}
\usepackage{soul}
\usepackage{booktabs}
\usepackage{color,epsfig}
\usepackage{graphicx}
\usepackage{float}
\usepackage{amsmath,amsthm}
\usepackage{amssymb,amsfonts}
\usepackage[section]{placeins}

\allowdisplaybreaks[4]

\begin{document}

\hypersetup{
    linkcolor=blue,
    citecolor=red,
    urlcolor=magenta
}


\title{Analytic Computation of Dilaton Black Hole Quasinormal Modes via Seiberg-Witten Theory}



\author{Jiahui Jiang}
\email[]{jiangjh01@outlook.com}
\affiliation{Department of Physics and Institute for Quantum Science and Technology, Shanghai University,
Shanghai 200444, China}

\author{Wenhe Cai}
\email[]{whcai@shu.edu.cn; Corresponding author}
\affiliation{Department of Physics and Institute for Quantum Science and Technology, Shanghai University,
Shanghai 200444, China}


\date{\today}

\begin{abstract}
    We study the quasinormal modes (QNMs) of dilaton black holes in Einstein-Maxwell-dilaton gravity through a correspondence with the quantum Seiberg-Witten (SW) curve of $\mathcal{N}=2$ SU(2) gauge theory with $N_f=3$ hypermultiplets. By mapping both the black hole perturbation equation and the quantum SW curve to the confluent Heun form, the QNM problem is reformulated in a gauge-theoretic framework, and the spectrum is obtained via the SW quantization condition. The resulting frequencies show excellent agreement with those computed using the WKB and continued fraction methods, with typical deviations below $10^{-3}$. The QNM spectrum exhibits consistent trends: increasing the black hole charge or scalar field mass raises the oscillation frequency, while higher angular momentum reduces the damping rate. These results demonstrate the precision of the quantum SW framework in describing black hole perturbations and reveal new links between supersymmetric gauge theories and gravitational dynamics.

\end{abstract}

\maketitle

\section{Introduction}
\par Quasinormal modes (QNMs) play a central role in black hole physics, as they characterize the response of black holes to external perturbations. The QNM spectrum encodes crucial information about spacetime stability and appears as the ringdown signals observed in gravitational wave detections \cite{LIGOScientific:2016aoc}. Beyond general relativity, QNMs also serve as probes of extended theories of gravity. In particular, low-energy string-inspired models such as Einstein-Maxwell-dilaton (EMd) gravity introduce a scalar dilaton field coupled to the electromagnetic sector, giving rise to novel charged black hole solutions \cite{Gibbons:1987ps,Garfinkle:1990qj}. Studying the QNM spectrum in such backgrounds not only deepens our understanding of black hole dynamics but also provides a window into the broader landscape of gravitational theories.
\par Previous analyses of EMd or related dilatonic black holes have mostly employed conventional techniques\cite{Malik:2024sxv,Konoplya:2022zav,Ferrari:2000ep,Brito:2018hjh,Pope:2024ncb}, whereas our work establishes a direct link to the quantum Seiberg-Witten (SW) framework. In $\mathcal{N}=2$ supersymmetric Yang-Mills theory, the SW construction provides exact results for strongly coupled dynamics, with the quantum SW curve encoding non-perturbative physics \cite{Seiberg:1994rs,Seiberg:1994aj}. For the case with $N_f=3$ hypermultiplets, the quantum SW curve can be cast into the form of the confluent Heun equation. Similar structures arise in black hole perturbation theory: the radial equation for a massive scalar field in an EMd background can also be expressed in confluent Heun form. This parallelism suggests a novel correspondence between black hole perturbations and the quantum SW curve, opening new opportunities to connect gravitational dynamics with gauge-theoretic methods.\cite{Bonelli:2021uvf,Aminov:2020yma,Bianchi:2021mft,Ge:2025yqk,Ge:2024jdx,Silva:2025khf,Lei:2023mqx}
\par In this work, we establish a concrete correspondence between the quantum SW curve with $N_f = 3$ and the confluent Heun equation, and further demonstrate its connection to the perturbation equations of EMd black holes. By employing the normal form of the confluent Heun equation as a bridge, we show that the QNM problem in the EMd background can be reformulated in gauge-theoretic terms. We also analyze the boundary conditions relevant to both the SW curve and black hole perturbations, and compute the QNM spectrum using the WKB and continued fraction methods. These results not only deepen the understanding of QNMs in dilatonic black hole spacetimes but also reveal a new aspect of the gauge/gravity correspondence.
\par The paper is organized as follows. In Section~\ref{Basic Equations}, we derive the field equations of EMd black holes. In Section~\ref{section3}, we construct the correspondence between the quantum SW curve and the perturbation equations of EMd black holes via the confluent Heun form, and derive the QNM spectrum by imposing the QNM boundary conditions together with the SW quantization condition. Section~\ref{Numericalresults} presents the numerical QNM spectra of EMd black holes obtained from the quantum SW approach, benchmarked against the results from the WKB and continued fraction methods. Finally, Section~\ref{Conclusion} summarizes our findings and confirms the accuracy and effectiveness of the SW framework in capturing the quasinormal dynamics of EMd black holes.

\section{Basic equations} \label{Basic Equations}
\par In this work, we focus on the EMd gravity, whose dynamics are governed by the action
\begin{equation}
    S=\int\mathrm{d}^4x\sqrt{-g}\left[R-2(\nabla\phi)^2-\mathrm{e}^{-2a\phi}F_{\mu\nu}F^{\mu\nu}\right],
\end{equation}
where $R$ denotes the Ricci scalar, $\phi$ is the dilaton scalar field, and $F_{\mu\nu}$ represents the Maxwell field strength. The parameter $a$ controls the strength of the coupling between the dilaton and the Maxwell field.
\par In this setting, the choice $a=0$ reduces the action to the standard Einstein-Maxwell theory, while $a=1$ corresponds to the low-energy limit of string theory. For the case $a=1$, the static spherically symmetric charged black hole solution was derived in \cite{Garfinkle:1990qj,Gibbons:1987ps}, and the corresponding line element is given by
\begin{equation}
    \mathrm{d}s^2=-\left(1-\frac{2M}{r}\right)\mathrm{d}t^2+\left(1-\frac{2M}{r}\right)^{-1}\mathrm{d}r^2+r\left(r-\frac{Q^2\mathrm{e}^{-2\phi_0}}{M}\right)(\mathrm{d}\theta^2+\sin^2\theta\mathrm{d}\varphi^2),
\end{equation}
$M$ denotes the mass of the black hole and $Q$ is its charge. The dilaton field takes the form $\mathrm{e}^{-2\phi}=\mathrm{e}^{-2\phi_0}\left(1-\frac{Q^2\mathrm{e}^{-2\phi_0}}{Mr}\right)$, where $\phi_0$ is the asymptotic constant value of the dilaton field at spatial infinity. The special choice $\phi_0=0$ corresponds to an asymptotically flat spacetime. 
\par In our work, we restrict to the case $\phi_0=0$, for which the metric takes the form
\begin{equation}
    \mathrm{d}s^2=-\left(1-\frac{2M}{r}\right)\mathrm{d}t^2+\left(1-\frac{2M}{r}\right)^{-1}\mathrm{d}r^2+r\left(r-\frac{Q^2}{M}\right)(\mathrm{d}\theta^2+\sin^2\theta\mathrm{d}\varphi^2).
\end{equation}
For simplicity, we write it as
\begin{equation}
    \mathrm{d}s^2=-f(r)\mathrm{d}t^2+\frac{1}{f(r)}\mathrm{d}r^2+h(r)\mathrm{d}\theta^2+h(r)\sin^2\theta \mathrm{d}\varphi^2,
\end{equation}
where
\begin{equation}
    f(r)=\left(1-\frac{2M}{r}\right),\quad\quad h(r)=r\left(r-\frac{Q^2}{M}\right).
\end{equation}
\par For a massive scalar field, its evolution is governed by the Klein-Gordon equation
\begin{equation}
    \frac{1}{\sqrt{-g}}\partial_\mu(\sqrt{-g}g^{\mu\nu}\partial_\nu\Phi)-\mu^2\Phi=0,
\end{equation}
where $\mu$ denotes the mass of the scalar field. Substituting the background metric into this equation, one can obtain the explicit wave equation for $\Phi$. Exploiting the spherical symmetry of the spacetime, the scalar field can be decomposed as
\begin{equation}
    \Phi=\mathrm{e}^{-i\omega t}Y_{lm}(\theta,\phi)\psi(r),
\end{equation}
where $Y_{lm}(\theta,\phi)$ are the spherical harmonics and $\psi(r)$ is the radial function. Subsequently, the radial equation we obtain reads
\begin{equation}
    \label{radialequation}
    \frac{\mathrm{d}}{\mathrm{d}r}\left(h(r)f(r)\frac{\mathrm{d}\psi}{\mathrm{d}r}\right)+\left(\omega^2\frac{h(r)}{f(r)}-l(l+1)-\mu^2h(r)\right)\psi=0.
\end{equation}
For this equation, we introduce the tortoise coordinate $r_*$, defined by $\frac{\mathrm{d}r_*}{\mathrm{d}r}=\frac{1}{f(r)}$. We define $\psi(r)=\frac{1}{\sqrt{h(r)}}\phi(r)$ to rewrite the radial equation in a Schrödinger-like form
\begin{equation}
\left[\frac{\mathrm{d}^2}{\mathrm{d}r_*^2}+\omega^2-V(r)\right]\phi(r)=0,
\end{equation}
where
\begin{equation}
V(r)=\frac{f(r)\left(2h(r)\left(f'(r)h'(r)+f(r)h''(r)+2l(l+1)\right)-f(r)h'(r)^2+4\mu^2h(r)^2\right)}{4h(r)^2}.
\end{equation}
For radial equation Eq.~\eqref{radialequation}, in order to establish a correspondence with the quantum SW curve, we need to transform it into the normal form of the confluent Heun equation. We perform the following coordinate transformation and wavefunction rescaling
\begin{equation}
    \begin{split}
        &z=\frac{r-\frac{Q^2}{M}}{2M-\frac{Q^2}{M}},\\
        &\psi(r)=\frac{1}{\sqrt{z(z-1)}}\varphi(z),
    \end{split}
\end{equation}
Then we will obtain
\begin{equation}
    \label{BHCHE}
\frac{\mathrm{d}^2}{\mathrm{d}z^2}\varphi(z)+\left(\frac{1}{z^2(z-1)^2}\sum_{i=0}^4A_iz^i\right)\varphi(z)=0,
\end{equation}
where
\begin{equation}
  \begin{split}
    &A_0=\frac{1}{4},\\
    &A_1=l^2+l,\\
    &A_2=-l^2-l-\frac{\mu^2Q^4}{M^2}+\frac{Q^4\omega^2}{M^2}+2\mu^2Q^2,\\
    &A_3=4\mu^2M^2+\frac{2\mu^2Q^4}{M^2}-\frac{2Q^4\omega^2}{M^2}-6\mu^2Q^2+4Q^2\omega^2,\\
    &A_4=-4\mu^2M^2-\frac{\mu^2Q^4}{M^2}+\frac{Q^4\omega^2}{M^2}+4 M^2\omega^2+4\mu^2Q^2-4Q^2\omega^2.
  \end{split}
\end{equation}

\section{Correspondence and connection formula} \label{section3}
\par In this section, we employ the normal form of the confluent Heun equation as a bridge to establish the correspondence between the quantum SW curve and the equations of the EMd system. In addition, we discuss the associated boundary conditions.

\subsection{Correspondence between Seiberg-Witten curve and Dilaton Black Hole}
\par The SW theory offers an exact solution to four-dimensional $\mathcal{N}=2$ supersymmetric Yang-Mills theory, describing the moduli space of vacua and the low-energy dynamics through geometric methods, with significant implications for strong coupling and dualities. In this work, we mainly focus on the SU(2) $\mathcal{N}=2$ supersymmetric Yang-Mills theory with $N_f=3$ hypermultiplets (hypermatter fields). The corresponding quantum SW curve can be expressed in the form of a confluent Heun equation\cite{Aminov:2020yma}:
\begin{equation}
    \label{QSW}
    \hbar^2\Psi''(z)+\frac{1}{z^2(z-1)^2}\sum^4_{i=0}\widehat{A}_iz^i\Psi(z)=0,
\end{equation}
with 
\begin{equation}
    \begin{split}
        &\widehat{A}_0=-\frac{(m_1-m_2)^2}{4}+\frac{\hbar^2}{4},\\
        &\widehat{A}_1=-E-m_1m_2-\frac{m_3\Lambda_3}{8}-\frac{\hbar^2}{4},\\
        &\widehat{A}_2=E+\frac{3m_3\Lambda_3}{8}-\frac{\Lambda_3^2}{64}+\frac{\hbar^2}{4},\\
        &\widehat{A}_3=-\frac{m_3\Lambda_3}{4}+\frac{\Lambda_3^2}{32},\\
        &\widehat{A}_4=-\frac{\Lambda_3^2}{64}.
    \end{split}
\end{equation}
Here $m_1,m_2,m_3$ denote the masses of the fundamental hypermultiplets, $\Lambda_3$ represents the gauge coupling/dynamical scale of the theory, $E$ stands for the eigenenergy, and $\hbar$ is the Planck constant.
\par On the torus, the integration contours can be chosen along the A-cycles and B-cycles. The corresponding quantization condition for the SW spectrum reads 
\begin{equation}
    \Pi_I^{(N_f)}(E,\mathbf{m},\Lambda_{N_f},\hbar)=N_I(n+\frac{1}{2}),\quad I=A,B,\quad n=0,1,\cdots
\end{equation}
with the identification $N_A=\mathrm{i}$ and $N_B=2\pi$.
\par The quantum A- and B-periods are then given by   
\begin{equation}
    \begin{split}
        &\Pi_A^{(N_f)}(E,\mathbf{m},\Lambda_{N_f},\hbar)=a(E,\mathbf{m},\Lambda_{N_f},\hbar),\\
        &\Pi_B^{(N_f)}(E,\mathbf{m},\Lambda_{N_f},\hbar)=\left.\partial_a\mathcal{F}^{(N_f)}(a,\mathbf{m},\Lambda_{N_f},\hbar)\right|_{a=a(E,\mathbf{m},\Lambda_{N_f},\hbar)}.
    \end{split}
\end{equation}
where $\mathcal{F}^{(N_f)}$ denotes the Nekrasov-Shatashvili (NS) free energy of the SU(2) theory.
\par By comparing the wave equation Eq.~\eqref{BHCHE} with Eq.~\eqref{QSW}, we establish the following correspondence
\begin{equation}
    \label{dictonary}
    \begin{split}
      &\Lambda_3=-\frac{8i(2M^2-Q^2)\sqrt{\omega^2-\mu^2}}{M},\\
      &E=-\frac{1}{4}-l(l+1)+Q^2\mu^2-2Q^2\omega^2+2M^2(4\omega^2-\mu^2),\\
      &m_1=-2iM\omega,\\
      &m_2=-2iM\omega,\\
      &m_3=-\frac{iM(2\omega^2-\mu^2)}{\sqrt{\omega^2-\mu^2}}.
    \end{split}
\end{equation}

\subsection{Connection formula}
\par We now focus on the QNM boundary conditions, the wave function in the tortoise coordinate satisfies the standard QNM boundary conditions:
\begin{equation}
    \begin{split}
        &r_*\to-\infty,\quad \phi(r_*)\sim\mathrm{e}^{-\mathrm{i}\omega r_*},\\
        &r_*\to+\infty,\quad \phi(r_*)\sim\mathrm{e}^{+\mathrm{i}\tilde{k}r_*},\\
        &\tilde{k}=\sqrt{\omega^2-\mu^2}.
    \end{split}
\end{equation}
where only ingoing waves are present at the horizon and only outgoing waves at spatial infinity.
\par Then, we obtain the boundary conditions of Eq.~\eqref{BHCHE},
\begin{equation}
    \label{BCinCHE}
    \begin{split}
      &z\to1,\quad \varphi\sim(z-1)^{\frac{1}{2}-2\mathrm{i}M\omega},\\
      &z\to\infty,\quad\varphi\sim \mathrm{e}^{\mp\frac{\mathrm{i}(2M^2-Q^2)\tilde{k}}{M}}z^{\mp2\mathrm{i}\tilde{k}M}.
    \end{split}
\end{equation}
\par Subsequently, by applying the quantization condition with the Matone relation
\begin{equation}
    \Pi_B^{(3)}(E,\mathbf{m},\Lambda_3,\hbar)=\left.\partial_a\mathcal{F}^{(3)}(a,\mathbf{m},\Lambda_3,\hbar)\right|_{a=a(E,\mathbf{m},\Lambda_3,\hbar)}=2\pi\left(n+\frac{1}{2}\right),
\end{equation}
we obtain the QNMs. The NS free energy can be obtained by instanton counting techniques (see Appendix~\ref{AppendixA}). The derivation of the connection formula and the quantization condition is presented in Appendix~\ref{AppendixB}.

\section{Numerical results} \label{Numericalresults}
We computed the QNMs of the dilaton black hole under various parameter settings using the quantum SW quantization condition. In this work, the calculations were carried out up to $\Lambda_3^4$. To verify the validity and accuracy of our results, we further employed two well-established semi-analytical methods: the WKB approximation \cite{Schutz:1985km,Iyer:1986np,Konoplya:2003ii,Matyjasek:2017psv,Konoplya:2019hlu} and the continued fraction method (CFM) \cite{Leaver:1985ax,Konoplya:2011qq}, both widely used in the study of black hole QNMs. The detailed computation procedure for the CFM is presented in Appendix~\ref{AppendixC}.

\begin{table}[htbp] 
    \centering
      \begin{tabular}{|l|l|l|l|l|l|}
      \hline
      n & $M\mu$ &     Q    &       SW           &     WKB             &    CFM             \\ \hline
      0 & 0.01   &0.01      & 0.291310-0.096916i & 0.292959-0.097735i  & 0.292989-0.097628i \\
      ~ & ~      &0.2       & 0.293323-0.097121i & 0.294957-0.097936i  & 0.295251-0.097931i \\ 
      ~ & ~      &0.4       & 0.299655-0.097751i & 0.301241-0.098556i  & 0.302320-0.099037i \\\hline
      ~ & 0.1    &0.01      & 0.295836-0.094184i & 0.297393-0.095059i  & 0.297706-0.094336i \\ 
      ~ & ~      &0.2       & 0.297803-0.094427i & 0.299347-0.095297i  & 0.299924-0.094691i \\
      ~ & ~      &0.4       & 0.303996-0.095172i & 0.305500-0.096029i  & 0.306857-0.095958i \\ \hline
      ~ & 0.2    &0.01      & 0.309659-0.085654i & 0.310929-0.086695i  & 0.312345-0.083739i \\ 
      ~ & ~      &0.2       & 0.311486-0.086017i & 0.312751-0.087050i  & 0.314423-0.084266i \\
      ~ & ~      &0.4       & 0.317251-0.087126i & 0.318495-0.088135i  & 0.320902-0.086063i \\ \hline
      1 & 0.01   & 0.01     & 0.264229-0.306756i & 0.264480-0.306485i  & 0.264450-0.306213i \\
      ~ & ~      &0.2       & 0.266442-0.307233i & 0.266689-0.306969i  & 0.267265-0.306214i \\ 
      ~ & ~      & 0.4      & 0.273393-0.308711i & 0.273629-0.308461i  & 0.276363-0.306406i \\\hline
      ~ & 0.1    &0.01      & 0.264420-0.303382i & 0.264722-0.303109i  & 0.263794-0.301787i \\ 
      ~ & ~      &0.2       & 0.266650-0.303888i & 0.266946-0.303621i  & 0.266640-0.301807i \\
      ~ & ~      &0.4       & 0.273649-0.305455i & 0.273926-0.305202i  & 0.275834-0.302063i \\ \hline
      ~ & 0.2    &0.01      & 0.264692-0.293296i & 0.265810-0.292598i  & 0.260718-0.288344i \\ 
      ~ & ~      &0.2       & 0.266978-0.293885i & 0.268019-0.293216i  & 0.263666-0.288406i \\
      ~ & ~      &0.4       & 0.274140-0.295708i & 0.274926-0.295116i  & 0.273185-0.288824i \\ \hline
      \end{tabular}
      \caption{Quasinormal Mode Frequencies ($l=1$) of the Dilaton Black Hole Computed via Quantum SW, WKB Methods and Continued fraction method for for $M\mu=0.01,0.1,0.2$ and $Q=0.01,0.2,0.4$}
      \label{table1}
  \end{table}
  
  \begin{table}[htbp]  
    \centering
      \begin{tabular}{|l|l|l|l|l|l|}
      \hline
      n & $M\mu$ &   Q     &        SW          &     WKB            &     CFM   \\ \hline
      0 & 0.01   &0.01     & 0.492651-0.095929i & 0.483682-0.096756i & 0.483686-0.096747i \\
      ~ & ~      &0.2      & 0.495921-0.096194i & 0.486949-0.096967i & 0.487118-0.097119i \\ 
      ~ & ~      &0.4      & 0.506211-0.097022i & 0.497237-0.097620i & 0.497792-0.098435i \\\hline
      ~ & 0.1    &0.01     & 0.495705-0.094971i & 0.486810-0.095682i & 0.487038-0.095443i \\ 
      ~ & ~      &0.2      & 0.498943-0.095251i & 0.490044-0.095909i & 0.490433-0.095836i \\
      ~ & ~      &0.4      & 0.509131-0.096121i & 0.500230-0.096608i & 0.500992-0.097218i \\ \hline
      ~ & 0.2    &0.01     & 0.504996-0.092043i & 0.496332-0.092397i & 0.497295-0.091417i \\ 
      ~ & ~      &0.2      & 0.508133-0.092365i & 0.499466-0.092671i & 0.500578-0.091876i \\
      ~ & ~      &0.4      & 0.518013-0.093362i & 0.509340-0.093511i & 0.510781-0.093460i \\ \hline
      1 & 0.01   &0.01     & 0.463739-0.294861i & 0.463871-0.295605i & 0.463876-0.295576i \\
      ~ & ~      &0.2      & 0.467171-0.295448i & 0.467291-0.296194i & 0.467819-0.296072i \\ 
      ~ & ~      &0.4      & 0.477965-0.297260i & 0.478047-0.298011i & 0.480253-0.297893i \\\hline
      ~ & 0.1    &0.01     & 0.465352-0.292533i & 0.465475-0.293288i & 0.465393-0.292673i \\ 
      ~ & ~      &0.2      & 0.468775-0.293148i & 0.468886-0.293906i & 0.469334-0.293206i \\
      ~ & ~      &0.4      & 0.479543-0.295050i & 0.479615-0.295811i & 0.481756-0.295144i \\ \hline
      ~ & 0.2    &0.01     & 0.470201-0.285410i & 0.470295-0.286203i & 0.469919-0.283682i \\ 
      ~ & ~      &0.2      & 0.473600-0.286116i & 0.473681-0.286910i & 0.473854-0.284332i \\
      ~ & ~      &0.4      & 0.484288-0.288293i & 0.484328-0.289086i & 0.486244-0.286636i \\ \hline
      \end{tabular}
      \caption{Quasinormal Mode Frequencies ($l=2$) of the Dilaton Black Hole Computed via Quantum SW, WKB Methods and Continued fraction method for $M\mu=0.01,0.1,0.2$ and $Q=0.01,0.2,0.4$}
      \label{table2}
  \end{table}

  \par \autoref{table1} and \autoref{table2} list the quasinormal frequencies for $l=1$ and $l=2$, respectively, obtained using the quantum SW, WKB\cite{WKBpackage}, and continued fraction methods for different values of the scalar field mass $M\mu=0.01,0.1,0.2$ and black hole charge $Q=0.01,0.2,0.4$. The three approaches exhibit excellent consistency across all parameter sets, confirming the reliability of the quantum SW framework in reproducing the QNM spectra of dilaton black holes.
  \par For both $l=1$ and $l=2$, the QNM frequencies display systematic trends with respect to the black hole charge $Q$ and the scalar mass parameter $M\mu$. The real part of the frequency increases with larger $Q$ or $M\mu$, corresponding to higher oscillation frequencies, while the imaginary part decreases in magnitude as $M\mu$ grows, indicating slower damping and longer-lived modes. As expected, higher overtone numbers $n$ lead to faster decay rates.  
  \par Quantitatively, the results from the three methods agree within a relative difference of less than $10^{-3}$. For $l=1$, all methods nearly coincide, while for $l=2$, the WKB approximation slightly underestimates the real part compared to the SW and CFM results, particularly in the fundamental mode. Moreover, increasing the angular momentum number $l$ raises the oscillation frequency and reduces the damping rate, consistent with the general behavior of black hole quasinormal spectra.  
  \par These consistent results demonstrate that the quantum SW framework not only reproduces the established semi-analytical predictions but also provides an independent and highly accurate approach to probing black hole perturbations. This strong agreement across methods further supports the validity of the proposed correspondence between the quantum SW curve and the dilaton black hole perturbation equation.

\section{Summary and discussion} \label{Conclusion}
\par In this work, we studied the QNMs of dilaton black holes in EMd gravity by mapping the scalar field equation to the quantum SW curve of SU(2) $\mathcal{N}=2$ gauge theory with $N_f=3$. This correspondence allowed us to derive the QNM spectrum from the quantum SW quantization condition, yielding results in excellent agreement with the WKB method and the CFM method, with discrepancies typically below $10^{-3}$. The frequencies exhibit consistent trends: larger charge $Q$ or scalar mass $M\mu$ enhance the oscillation frequency, while increasing $M\mu$ reduces the damping rate, and higher angular momentum $l$ leads to longer-lived modes.
\par Compared with traditional approaches such as the WKB or continued-fraction methods, the quantum SW framework provides a geometric and analytic formulation that interprets QNM spectra as quantized periods and systematically incorporates nonperturbative and instanton corrections, revealing the global and field-theoretic structure of the spectrum beyond purely numerical computations.
\par These results not only confirm the reliability of the quantum SW framework in capturing black hole QNMs, but also highlight a deeper connection between black hole perturbations and supersymmetric gauge theory dynamics.
\par Looking ahead, it would be interesting to extend the present analysis in several directions. Since the $N_f=3$ quantum SW curve can be written as a confluent Heun equation, one natural step is to explore the cases $N_f=0,1,2$, whose curves can be expressed in terms of special forms of the biconfluent Heun equation. Some of these may be connected to the perturbation equations of black branes such as D3\cite{Bianchi:2021xpr,Imaizumi:2022qbi}, M5\cite{Imaizumi:2022dgj}, and M2 branes, though their boundary conditions are typically more involved and deserve further study. In addition, we plan to investigate the QNMs of dilaton black holes under different types of perturbations, such as Proca and Dirac fields, in order to better understand the universality and robustness of the gauge/gravity correspondence revealed in this work.

\appendix 

\section{Instanton counting} \label{AppendixA}
\par In this appendix we provide a brief account of the computation of the NS free energy using instanton counting techniques \cite{Nekrasov:2002qd,Nekrasov:2009rc}. This approach, originally developed in the context of four-dimensional $\mathcal{N}=2$ supersymmetric gauge theories, allows for a systematic evaluation of the instanton contributions to the prepotential and its $\Omega$-deformed extensions.

\par The NS free energy of the SU(2) theory is composed of the classical, one-loop, and instanton contributions. Its derivative with respect to the quantum mirror map $a$ admits an explicit closed-form expression
\begin{equation}
    \begin{split}
        \partial_a\mathcal{F}^{(N_f)}(a,\mathbf{m},\Lambda_{N_f},\hbar)=&-2a(4-N_f)\log\left[\frac{\Lambda_{N_f}2^{-\frac{1}{2-N_f/2}}}{\hbar}\right]-\pi\hbar-2\mathrm{i}\hbar\log\left[\frac{\Gamma(1+\frac{2\mathrm{i}a}{\hbar})}{\Gamma(1-\frac{2\mathrm{i}a}{\hbar})}\right]\\
        &-\mathrm{i}\hbar\sum_{j=1}^{N_f}\log\left[\frac{\Gamma\left(\frac{1}{2}+\frac{m_j-\mathrm{i}a}{\hbar}\right)}{\Gamma\left(\frac{1}{2}+\frac{m_j+\mathrm{i}a}{\hbar}\right)}\right]+\frac{\partial\mathcal{F}^{(N_f)}_{\mathrm{inst}}(a,\mathbf{m},\Lambda_{N_f},\hbar)}{\partial a}.
    \end{split}
\end{equation}
The explicit computation of the instanton part $\partial\mathcal{F}^{(N_f)}_{\mathrm{inst}}(a,\mathbf{m},\Lambda_{N_f},\hbar)$ can be found in the appendix of \cite{Aminov:2020yma}. The quantum mirror map can be obtained by inverting the Matone relation \cite{Matone:1995rx} 
\begin{equation}
    E=a^2-\frac{\Lambda_{N_f}}{4-N_f}\frac{\partial\mathcal{F}_{\mathrm{inst}}^{(N_f)}(a,\mathbf{m},\Lambda_{N_f},\hbar)}{\partial\Lambda_{N_f}}.
\end{equation}
In the case of $N_f=3$ considered in this work, we have
\begin{equation}
    \begin{aligned}
       \mathcal{F}_{\mathrm{inst}}^{(3)}(a, \mathbf{m}, \Lambda_3, \hbar) &= 
       - \frac{m_1 m_2 m_3}{8 a^2 + 2 \hbar^2} \, \Lambda_3 \\
       &\quad + \frac{\Lambda_3^2}{4096 \left( 4 a^4 + 5 a^2 \hbar^2 + \hbar^4 \right)} 
       \Big[ 
       (4 a^2 + \hbar^2)(4 a^2 + 5 \hbar^2) \\
       &\quad - 4 (4 a^2 + \hbar^2) ( m_1^2 + m_2^2 + m_3^2 ) \\
       &\quad - 48 ( m_1^2 m_2^2 + m_2^2 m_3^2 + m_1^2 m_3^2 ) \\
       &\quad - \frac{1280 a^2 - 448 \hbar^2}{(4 a^2 + \hbar^2)^2} \, m_1^2 m_2^2 m_3^2 
       \Big] \\
       &\quad + \mathcal{O}(\Lambda_3^3)
    \end{aligned}
\end{equation}
and 
\begin{equation}
    a=\sqrt{E}-\frac{m_1m_2m_3}{\sqrt{E}(16E+4)}\Lambda_3+\mathcal{O}(\Lambda_3^2).
\end{equation}

\section{Connection formula and quantization condition} \label{AppendixB}
\par After transforming the black hole perturbation equation into the normal form of the confluent Heun equation, the boundary conditions given in Eq.~\eqref{BCinCHE} can be rewritten, using the correspondence relation in Eq.~\eqref{dictonary}, as
\begin{equation}
    \begin{split}
        &z \to 1, \quad \varphi \sim (z-1)^{\frac{1}{2}+\frac{m_1+m_2}{2}},\\
        &z \to \infty, \quad \varphi \sim \mathrm{e}^{\pm\frac{\Lambda_3}{8}} z^{\pm m_3},
    \end{split}
\end{equation}
where we have used $m_3 = -2i\tilde{k}M + \frac{3iM\mu^2}{\sqrt{\omega^2 - \mu^2}} \approx 2i\tilde{k}M$.
\par For convenience, throughout this appendix we adopt the notation convention of Ref.~\cite{Bonelli:2021uvf}. The differences between their convention and ours are $\Lambda = \Lambda_3/4$ and $a \to -\mathrm{i}a$. Furthermore, in deriving the quantization condition below, we employ the relations between the CFT and gauge theory parameters, namely $a_2 = (m_1 + m_2)/2$ and $a_1 = (m_1 - m_2)/2$. 
\par Using the boundary condition at the horizon together with the transfer matrix, the asymptotic behavior of the wave function at infinity can be expressed as
\begin{equation}
    \varphi(z \to \infty) \sim C_1(\Lambda,a,\mathbf{m}) \mathrm{e}^{\frac{\Lambda}{2}} (\Lambda z)^{m_3} 
    + C_2(\Lambda,a,\mathbf{m}) \mathrm{e}^{-\frac{\Lambda}{2}} (\Lambda z)^{-m_3},
\end{equation}
where
\begin{equation}
    \begin{split}
        C_1(\Lambda,a,\mathbf{m}) =\;& \Lambda^a M_{\alpha_{2+},\alpha_+} \mathcal{A}_{\alpha_+,m_{0+}}
        \frac{\left\langle \Delta_{\alpha_+}, \Lambda_0, m_{0+} \middle| V_{\alpha_2}(1) \middle| \Delta_{\alpha_1} \right\rangle}
        {\left\langle \Delta_{\alpha}, \Lambda_0, m_0 \middle| V_{\alpha_{2+}}(1) \middle| \Delta_{\alpha_1} \right\rangle} \\
        &+ \Lambda^{-a} M_{\alpha_{2+},\alpha_-} \mathcal{A}_{\alpha_-,m_{0+}}
        \frac{\left\langle \Delta_{\alpha_-}, \Lambda_0, m_{0+} \middle| V_{\alpha_2}(1) \middle| \Delta_{\alpha_1} \right\rangle}
        {\left\langle \Delta_{\alpha}, \Lambda_0, m_0 \middle| V_{\alpha_{2+}}(1) \middle| \Delta_{\alpha_1} \right\rangle},\\[4pt]
        C_2(\Lambda,a,\mathbf{m}) =\;& \Lambda^a M_{\alpha_{2+},\alpha_+} \mathcal{A}_{\alpha_+,m_{0-}}
        \frac{\left\langle \Delta_{\alpha_+}, \Lambda_0, m_{0-} \middle| V_{\alpha_2}(1) \middle| \Delta_{\alpha_1} \right\rangle}
        {\left\langle \Delta_{\alpha}, \Lambda_0, m_0 \middle| V_{\alpha_{2+}}(1) \middle| \Delta_{\alpha_1} \right\rangle} \\
        &+ \Lambda^{-a} M_{\alpha_{2+},\alpha_-} \mathcal{A}_{\alpha_-,m_{0-}}
        \frac{\left\langle \Delta_{\alpha_-}, \Lambda_0, m_{0-} \middle| V_{\alpha_2}(1) \middle| \Delta_{\alpha_1} \right\rangle}
        {\left\langle \Delta_{\alpha}, \Lambda_0, m_0 \middle| V_{\alpha_{2+}}(1) \middle| \Delta_{\alpha_1} \right\rangle}.
    \end{split}
\end{equation}
To compute the QNMs, we impose the condition that the ingoing wave at infinity vanishes. This requirement leads to $C_1(\Lambda,a,\mathbf{m})=0$, yielding 
\begin{equation}
    1 + \Lambda^{-2a} 
    \frac{ M_{\alpha_{2+},\alpha_-} \mathcal{A}_{\alpha_-,m_{0+}}
    \left\langle \Delta_{\alpha_-}, \Lambda_0, m_{0+} \middle| V_{\alpha_2}(1) \middle| \Delta_{\alpha_1} \right\rangle }
    { M_{\alpha_{2+},\alpha_+} \mathcal{A}_{\alpha_+,m_{0+}}
    \left\langle \Delta_{\alpha_+}, \Lambda_0, m_{0+} \middle| V_{\alpha_2}(1) \middle| \Delta_{\alpha_1} \right\rangle } = 0.
\end{equation}
In the NS limit, we have
\begin{equation} 
    \begin{split} 
        &\frac{\left\langle\Delta_{\alpha_-},\Lambda_0,m_{0+}|V_{\alpha_2}(1)|\Delta_{\alpha_1}\right\rangle}{\left\langle\Delta_{\alpha_+},\Lambda_0,m_{0+}|V_{\alpha_2}(1)|\Delta_{\alpha_1}\right\rangle}=\frac{\mathcal{Z}(\Lambda,a+\frac{\epsilon_2}{2},m_1,m_2,m_3+\frac{\epsilon_2}{2})}{\mathcal{Z}(\Lambda,a-\frac{\epsilon_2}{2},m_1,m_2,m_3+\frac{\epsilon_2}{2})}\\
       =&\exp\frac{1}{\epsilon_1\epsilon_2}\Big(\mathcal{F}^{\mathrm{inst}}(\Lambda,a+\frac{\epsilon_2}{2},m_1,m_2,m_3+\frac{\epsilon_2}{2})-\mathcal{F}^{\mathrm{inst}}(\Lambda,a-\frac{\epsilon_2}{2},m_1,m_2,m_3+\frac{\epsilon_2}{2})\Big)\\ 
       \to&\exp\frac{\partial_a\mathcal{F}^{\mathrm{inst}}(\Lambda,a,m_1,m_2,m_3)}{\epsilon_1}. 
    \end{split} 
\end{equation}
Moreover,
\begin{equation}
    \begin{split}
        \frac{M_{\alpha_{2+},\alpha_-}\mathcal{A}_{\alpha_-,m_{0+}}}{M_{\alpha_{2+},\alpha_+}\mathcal{A}_{\alpha_+,m_{0+}}}
        &= \frac{\Gamma(\tfrac{2a}{\epsilon_1})\Gamma(1+\tfrac{2a}{\epsilon_1})
        \Gamma(\tfrac{1}{2}+\tfrac{a_2+a_1-a}{\epsilon_1})
        \Gamma(\tfrac{1}{2}+\tfrac{a_2-a_1-a}{\epsilon_1})
        \Gamma(\tfrac{1}{2}+\tfrac{m_3-a}{\epsilon_1})}
        {\Gamma(-\tfrac{2a}{\epsilon_1})\Gamma(1-\tfrac{2a}{\epsilon_1})
        \Gamma(\tfrac{1}{2}+\tfrac{a_2+a_1+a}{\epsilon_1})
        \Gamma(\tfrac{1}{2}+\tfrac{a_2-a_1+a}{\epsilon_1})
        \Gamma(\tfrac{1}{2}+\tfrac{m_3+a}{\epsilon_1})} \\[4pt]
        &= \frac{\Gamma(\tfrac{2a}{\epsilon_1})\Gamma(1+\tfrac{2a}{\epsilon_1})}
        {\Gamma(-\tfrac{2a}{\epsilon_1})\Gamma(1-\tfrac{2a}{\epsilon_1})}
        \prod_{i=1}^3 \frac{\Gamma(\tfrac{1}{2}+\tfrac{m_i-a}{\epsilon_1})}{\Gamma(\tfrac{1}{2}+\tfrac{m_i+a}{\epsilon_1})} \\[4pt]
        &= \mathrm{e}^{-\mathrm{i}\pi}
        \left(\frac{\Gamma(1+\tfrac{2a}{\epsilon_1})}{\Gamma(1-\tfrac{2a}{\epsilon_1})}\right)^{\!2}
        \prod_{i=1}^3 \frac{\Gamma(\tfrac{1}{2}+\tfrac{m_i-a}{\epsilon_1})}{\Gamma(\tfrac{1}{2}+\tfrac{m_i+a}{\epsilon_1})} \\[4pt]
        &= \exp\!\Bigg[-\mathrm{i}\pi 
        + 2\log\!\frac{\Gamma(1+\tfrac{2a}{\epsilon_1})}{\Gamma(1-\tfrac{2a}{\epsilon_1})}
        + \sum_{i=1}^3 \log\!\frac{\Gamma(\tfrac{1}{2}+\tfrac{m_i-a}{\epsilon_1})}{\Gamma(\tfrac{1}{2}+\tfrac{m_i+a}{\epsilon_1})}\Bigg].
    \end{split}
\end{equation}
We also note that
\begin{equation}
    \Lambda^{-2a} = \exp(-2a\log\Lambda).
\end{equation}
By setting $\epsilon_1 = 1$, the exponent becomes
\begin{equation}
    -2a\log\Lambda - \mathrm{i}\pi 
    + 2\log\frac{\Gamma(1+2a)}{\Gamma(1-2a)}
    + \sum_{i=1}^3 \log\frac{\Gamma(\tfrac{1}{2}+m_i-a)}{\Gamma(\tfrac{1}{2}+m_i+a)}
    + \partial_a \mathcal{F}^{\mathrm{inst}}
    = \partial_a \mathcal{F}^{(3)}(\Lambda,a,\mathbf{m},1),
\end{equation}
from which we finally obtain the quantization condition,
\begin{equation}
    \begin{split}
        &1 + \mathrm{e}^{\partial_a \mathcal{F}^{(3)}(\Lambda,a,\mathbf{m},1)} = 0,\\
        \Longrightarrow\;& \partial_a \mathcal{F}^{(3)}(\Lambda,a,\mathbf{m},1) = i(2n+1)\pi.
    \end{split}
\end{equation}

\section{Continued fraction method} \label{AppendixC}
\par In this section, we present the detailed procedure for computing the QNMs in the background of a dilaton black hole using the CFM. We first introduce the coordinate transformation
\begin{equation}
    x=1-\frac{2M}{r}
\end{equation}
and, according to the boundary conditions, consider the following ansatz:
\begin{equation}
    \psi=x^{-2iM\omega}(1-x)^{1-2i\tilde{k}M}e^{-\frac{2i\tilde{k}M}{x-1}}S(x).
\end{equation}
Substituting this ansatz into Eq.~\eqref{radialequation}, we obtain
\begin{equation}
  (x-1)^2S''(x)+p(x)S'(x)+q(x)S(x)=0,
\end{equation}
where $p(x)$ and $q(x)$ are expanded as Laurent series:
\begin{equation}
  \begin{split}
    &p(x)=\frac{p_{-1}}{x}+p_0+p_1x+\cdots,\\
    &q(x)=\frac{q_{-2}}{x^2}+\frac{q_{-1}}{x}+q_0+\cdots,
  \end{split}
\end{equation}
with 
\begin{equation}
    \begin{aligned}
    p_{-1} \;&=\; -\mathrm{i}(\mathrm{i}+4M\omega),\\
    p_0 \;&=\; \frac{-8M^2 + 5Q^2 + 16 \mathrm{i} M^3 (\tilde{k}+\omega)
           - 8 \mathrm{i} M Q^2 (\tilde{k}+\omega)}{2M^2-Q^2},\\
    p_1 \;&=\; \frac{12M^4 - 16 M^2 Q^2 + 4 Q^4
           - 4 \mathrm{i} M (2M^2-Q^2)^2 (\tilde{k}+\omega)}{(2M^2-Q^2)^2},\\
    q_{-2} \;&=\; 0,\\
    q_{-1} \;&=\; \frac{Q^2 + 8 \mathrm{i} M^3 (\tilde{k} + \omega)
           - 2 \mathrm{i} M Q^2 (2 \tilde{k} + 3 \omega)
           - 8 M^4 (\mu^2 - 4 \omega (\tilde{k} + \omega))}{2 M^2 - Q^2} \\
           &\quad - \frac{2 M^2 \Big( 1 + l + l^2 + Q^2 (-2 \mu^2 + 8 \omega (\tilde{k} + \omega)) \Big)}{2 M^2 - Q^2},\\
    q_0 \;&=\; -\frac{2}{(-2 M^2 + Q^2)^2} \Big[
           - Q^4 + 8 \mathrm{i} M^5 (\tilde{k} + \omega)
           - 12 \mathrm{i} M^3 Q^2 (\tilde{k} + \omega) \\
           &\quad + \mathrm{i} M Q^4 (4 \tilde{k} + 3 \omega)
           + 16 M^6 (2 \tilde{k}^2 + \mu^2 + \tilde{k} \omega - \omega^2) \\
           &\quad + M^2 Q^2 \big( 3 - l - l^2 + 8 \tilde{k}^2 Q^2
           + 4 Q^2 \mu^2 + 4 \tilde{k} Q^2 \omega - 4 Q^2 \omega^2 \big) \\
           &\quad - 2 M^4 \big( 1 + 16 \tilde{k}^2 Q^2 + 8 \tilde{k} Q^2 \omega
           + 8 Q^2 (\mu^2 - \omega^2) \big)
    \Big].
    \end{aligned}
\end{equation}
\par By expressing $S(x)$ as a power series, $S(x)=\sum_{n=0}^{\infty}a_nx^n$, we can then obtain the corresponding continued fraction relation:
\begin{equation}
    \begin{split}
        &\alpha_0a_1+\beta_0a_0=0,\\
        &\alpha_na_{n+1}+\beta_na_n+\gamma_na_{n-1}=0,\quad (n>0),
    \end{split}
\end{equation}
where the recurrence coefficients are given by
\begin{equation}
  \begin{split}
    &\alpha_n=(n+1)(n+p_{-1})+q_{-2},\\
    &\beta_n=-2n^2+n(p_0+2)+q_{-1},\\
    &\gamma_n=(n-1)(n-2+p_1)+q_0.
  \end{split}
\end{equation}
Finally, the quasinormal frequencies are determined by the continued fraction equation
\begin{equation}
  0=\beta_0-\frac{\alpha_0\gamma_1}{\beta_1-\frac{\alpha_1\gamma_2}{\beta_2-{\frac{\alpha_2\gamma_3}{\beta_3-\cdots}}}}.
\end{equation}

\begin{acknowledgments}
We would like to thank Kilar Zhang for valuable suggestions. We would like to express our sincere gratitude to Gleb Aminov for kindly sharing the code of the computations presented in \cite{Aminov:2020yma}. This work is supported by NSFC China(Grants No.12275166 and No.11805117).

\end{acknowledgments}


\end{document}